\documentclass[twocolumn,nofootinbib,showpacs,epsfig,prl,reprint]{revtex4}

\usepackage{amsmath}
\usepackage{amsfonts}
\usepackage{amssymb}
\usepackage{graphicx}
\usepackage{color}
\usepackage{xcolor}

\newcommand{\be}{\begin{equation}}
\newcommand{\ee}{\end{equation}}
\newcommand{\reff}[1]{(\ref{#1})}
\newcommand{\suml}[2]{\sum\limits_{#1}^{#2}}
\newcommand{\sysb}{\left\{\begin{array}}
\newcommand{\syse}{\end{array}\right.}
\newcommand{\rmd}{{\rm{d}}}
\newcommand{\comm}[2]{\left[ #1, #2 \right]}

\newcommand{\lan}{\left\langle}
\newcommand{\ran}{\right\rangle}

\begin{document}

\title{Pre-thermalization in a non-integrable quantum spin chain after a quench}

\author{Matteo Marcuzzi${}^{\dag,\ddag}$\footnote{These two authors equally contributed to the work.}, Jamir Marino${}^{\dag,\ddag}$\footnotemark[1], Andrea Gambassi${}^{\dag,\ddag}$ and Alessandro Silva${}^{\dag,1}$}

\address{$^\dag$SISSA --- International School for Advanced Studies, via Bonomea 265, 34136 Trieste, Italy}
\address{$^\ddag$INFN --- Istituto Nazionale di Fisica Nucleare, sezione di Trieste}
\address{$^1$ICTP --- International Centre for Theoretical Physics, P.O. Box 586, 34014 Trieste, Italy}

\begin{abstract}
We study the dynamics of a quantum Ising chain after the sudden introduction of a non-integrable long-range interaction. 
Via an exact mapping onto a fully-connected lattice of hard-core bosons, we show that a pre-thermal state emerges and we investigate 
its features by focusing on a class of physically relevant observables.
In order to gain insight into the eventual thermalization, we outline a diagrammatic approach which complements the study of the previous quasi-stationary state and provides the basis for a  self-consistent solution of the kinetic equation.  This analysis suggests that both the temporal decay towards the pre-thermal state and the crossover to the eventual thermal one may occur algebraically.
\end{abstract}




\maketitle

In the past decade, the impressive progress in manipulating cold atomic gases spurred the
interest in the non-equilibrium dynamics of isolated, strongly interacting quantum many-body systems \cite{coldatoms}.
In particular, the observed lack of thermalization 
in one-dimensional (1D) Bose gases \cite{Greiner2002,Kinoshita2006} close to integrability generated an intense theoretical activity devoted to 
understanding thermalization in many-body systems after a sudden change (\emph{quench}) of a control parameter.
Despite the conjectured relations between the absence of thermalization and the integrability of a system \cite{IntegrvsTh}, as well as between thermalization, quantum chaos~\cite{Sred}, the eigenstate thermalization hypothesis~\cite{Rigol2} and the localized/delocalized structure of many-body eigenstates~\cite{MBL}, 
the features of the approach to the eventual thermal state remain largely unexplored (see, e.g., Ref.~\cite{Polkovnikov2011}). 
The recent theoretical proposal~\cite{BergPret} and experimental observation of a two-stage relaxation involving a \emph{pre-thermal} quasi-stationary state call for a better understanding of the relaxation kinetics of isolated quantum many-body systems~\cite{pretherm}.\\
\indent In integrable quantum many-body systems, the existence of a maximal set of conserved quantities allows an exact analytical study of the relaxation 
towards a non-thermal Generalized Gibbs Ensemble (GGE), which maximizes the entropy under the constraint of fixed integrals of motion \cite{IntegrvsTh,Jaynes1957,Barthel}.  
If integrability is weakly broken (which is always the case in experiments) \cite{Kollar2011}, a many-body system initially prepared in the ground state of an integrable Hamiltonian does not directly thermalize but it may be trapped in an intermediate quasi-stationary state similar to the GGE of the integrable counterpart. The resulting  \emph{pre-thermalization} \cite{BergPret}  has been studied both theoretically in Fermi \cite{Moeckel2010} and Luttinger liquids \cite{Mitra}, in long-range quantum Ising models \cite{Kastner}, and experimentally in 1D quasi-condensates \cite{pretherm}.  Despite this progress, the description of the breaking of integrability is technically challenging and therefore restricted either to the study of small systems \cite{small} or to the numerical analysis of perturbative quantum kinetic equations~\cite{TavMit}.

In this work, we introduce a simple, prototypical model for studying thermalization and pre-thermalization:  
a quantum Ising chain (QIC) perturbed away from integrability by a long-range spin-spin interaction.
Even though many of the conservation laws of the QIC are violated by this long-range interaction, we show that the model 
maps into one of hard-core bosons hopping on a lattice. 
Within the latter, pre-thermalization occurs naturally for small quenches: as long as the quasi-particle density remains sufficiently low, the hard-core constraint is not effective and the model can be solved numerically up to a quite large size, distinctively showing pre-thermal plateaux in the dynamics of physically relevant observables.
The associated quasi-stationary values are typically approached algebraically in time. 
Features of the pre-thermal state are reproduced within a finite-order diagrammatic perturbation theory, while 
the departure from the pre-thermal state towards thermal equilibrium requires a kinetic equation derived within the self-consistent Born approximation. We show that the latter admits the thermal distribution as stationary state and that the spectrum of relaxation rates of weak perturbations around the equilibrium solution approaches zero continuously. 
The absence of a non-vanishing minimal relaxation rate
suggests that also the relaxation towards the thermal state occurs algebraically.

\emph{The model.---}We consider the effect of an integrability-breaking perturbation on the dynamics of the QIC~\cite{Sachdev}, 
\begin{equation}
\label{hamspin}
H_0(g)=-\suml{i=1}{N} \left( \sigma_i^x\sigma_{i+1}^x+g\sigma_i^z \right), 
\end{equation}
where $\sigma^{(x,y,z)}$ are Pauli matrices, $N$ is the total number of spins, and $g$ the strength of the transverse field. This model, which undergoes a prototypical quantum phase transition between a paramagnetic ($g>1$) and a ferromagnetic ($g<1$)  phase, has been extensively studied both in and out of equilibrium, taking advantage of its integrability. 
After a Jordan-Wigner transformation followed by a Bogoliubov rotation, $H_0$ becomes 
$H_0=\sum_{k>0}\epsilon_k\psi_k^\dag\sigma^z\psi_k$ in terms of spinors $\psi_k\equiv\bigl(\begin{smallmatrix}\gamma_k\\ \gamma^\dag_{-k}\end{smallmatrix} \bigr)$ where $\gamma_k$ are fermionic operators, and $\epsilon_k\equiv\sqrt{1+g^2-2 g\cos k}$ is the energy of the quasi-particles~\cite{Sachdev}. The dynamics after a quench of the transverse field $g_0 \rightarrow g$ has been thoroughly investigated both numerically 
and analytically (see, e.g., Refs.~\cite{Ising,Foini,Fagotti}).

Focusing for simplicity on quenches within the same phase (say $g_0,g>1$), the integrability of $H_0$ makes the dynamics trivial in the quasi-particle representation: the ground state $| 0 \rangle_{g_0}$ of $H_0(g_0)$, in fact,  can be represented as a BCS state $| 0 \rangle_{g_0} \propto \exp\{-\sum_k B_{g_0 \rightarrow g}(k) b^{\dagger}_k \} | 0 \rangle_{g}$ \cite{Essler}, in terms of the ground state $| 0 \rangle_{g}$ of $H_0(g)$ and of the \emph{pair} operators $b^{\dagger}_k = \gamma^{\dagger}_k\gamma^{\dagger}_{-k}$ of its quasi-particles $\gamma_k$. 
The initial distribution of these \emph{zero-momentum} fermionic pairs, determined by $B_{g_0 \rightarrow g}$  is not modified by the time evolution and therefore it affects the asymptotic values of local observables, described consequently by a GGE. 

The peculiar structure of the initial state and of the subsequent dynamics is generically spoiled by breaking the integrability of the model, which is expected to cause scattering not only of pairs, but also of individual quasi-particle modes $\gamma_k$; as a consequence, the energy initially injected into the system gets redistributed among the various modes, eventually leading to thermalization.
In order to make progress in understanding thermalization and pre-thermalization it is particularly valuable to have at hand a simple enough model in which the breaking of integrability is amenable to both analytic and numerical analysis in a controlled and physically transparent way.
Such an instance is provided by a quantum chain with Hamiltonian $H_0+V$,  where $H_0$ is given by Eq.~\reff{hamspin} and
\be
V=\frac{\lambda}{N} (M_z-\overline{M_z})^2,
\label{eq:V}
\ee
where $M_z=\sum_i\sigma_i^z$ is the global transverse magnetization and the operator $\overline{M_z}$ its long-time temporal average 
in the absence of perturbation $\lambda=0$. The subtraction in Eq.~\reff{eq:V} cancels the constants of motion $\hat n_k = \gamma_k^\dag \gamma_k$ present in the definition of $M_z$ (c.f., Eq.~\reff{hambog}) and is made in order to recover the temporal cluster property of the two-time correlation functions of the subtracted transverse magnetization $M_z-\overline{M_z}$ in the long-time limit \cite{Foini}.
In particular, we consider a quench from $H_0(g_0)$ at $t<0$ to $H \equiv H_0(g)+V$ at $t>0$.
$V$ effectively breaks the integrability of $H_0(g)$ in terms of its Bogoliubov fermions $\gamma_k$ and introduces scattering among \emph{zero-momentum pairs} in the fermionic representation.

\emph{The mapping.---}The spin chain described by $H$ can be conveniently mapped onto a quadratic (yet non-diagonal) Hamiltonian of hardcore bosons, as we show here.
First of all, we take advantage of the fermionic representation of $H_0(g_0)$ in order to better understand 
the effect of breaking the integrability. In fact, in terms of $\gamma_k$, $V$ becomes
\begin{equation}\label{hambog}
V= \frac{\lambda}{N}\Big[\sum_{k>0}\sin (2\theta_k) \; \psi_k^\dag\sigma^y\psi_k\Big]^2,
\end{equation}
where $\theta_k(g)$ is the Bogoliubov angle \cite{Sachdev} with $\tan (2\theta_k) = (\sin k)/(g-\cos k)$.
Note that $I_k = \hat n_k - \hat n_{-k}$ commutes with $H_0(g)$ for all $k>0$ and therefore $\{I_k\}_k$ is a set of $N/2$ constants of motion \cite{Fagotti}. 
The eigenvalues of $I_k$ are $0$ and $\pm 1$, corresponding to states in which two quasi-particles with momenta $\pm k$ are either simultaneously present or absent, and to states in which only one of the two is present, respectively. 
Accordingly, the configuration space is split in eigensectors characterized by the string of the $N/2$ possible eigenvalues of $\{I_k\}_k$, with dimension $2^{N_0}$, $N_0$ being the number of $0$s present in the corresponding string. 
These sectors are closed under the action of two-fermion operators such as  the number operator $\hat n_k$, and the pair creation $b_k^\dag$ and annihilation $b_k$ operators, in terms of which $H$ becomes  
\be
\sysb{l}
H = \suml{k>0}{} \left[\epsilon_k - (\lambda/N) \sin^2(2\theta_k)\right] \left( I^2_k - 1 \right) + H', \\
H' = \suml{k,q>0}{} \left[2\beta_{kq} b_k^\dag b_q - \alpha_{kq}  (b_k^\dag b_q^\dag + b_k b_q)\right],
\syse
\label{chernabog}
\ee
where $\alpha_{kq} = (\lambda/N) (1-\delta_{kq})\sin(2\theta_k) \sin(2\theta_q)$ and $\beta_{kq} = \epsilon_k \delta_{kq} + \alpha_{kq}$. The ``$b$'' operators commute at different momenta and anticommute at equal ones, except for $\{ b_k , b_k^\dag \} = 1 - I^2_k$, and thus they behave almost as hard-core bosons. On the other hand, it is useful to notice that, in a sector with $I_k = \pm 1$, both $b_k$ and $b_k^\dag$ act as the null operator and can be effectively expunged from $H'$, leaving behind only those corresponding to momenta $q$ for which $I_q = 0$, which can be treated instead as bona-fide hard-core bosons. Summarizing, within a sector characterized by having $N/2 - N_0$ unpaired quasi-particles,  
$H$ describes a fully-connected model of hard-core bosons on a lattice with $N_0$ sites. 

\emph{Prethermalization.---}This representation allows a consistent description of pre-thermalization and thermalization based on standard approximations. 
Note that, although $H'$ is quadratic in the pair operators, it cannot be trivially integrated via a Bogoliubov rotation, for that could not preserve the mixed (anti)commutation relations. 
Some progress can be made, instead, by re-expressing $H$ in terms of bosonic operators $a_k$ with $\comm{a_k}{a_q^\dag} = \delta_{kq}$, via a Holstein-Primakoff transformation \cite{HP} $b_k = (1-a^\dag_k a_k)^{1/2} \,a_k$, $b_k^\dag = a_k^\dag (1-a^\dag_k a_k)^{1/2}$.
Assuming a low density of excitations (i.e., a small quench), one can expand the square roots to lowest order $b_k \simeq a_k$, $b_k^\dag \simeq a_k^\dag$ and $H$ becomes an Hamiltonian of free bosons (whose dynamics describes pre-thermalization), while higher-order terms introduce the  interactions which are expected to lead to thermalization. For this approximation to be valid, the mean bosonic populations have to remain small during the evolution. As heuristically expected, this occurs the smaller the energy injected at $t=0$ is, i.e., the smaller the amplitude $|g-g_0|$ of the quench is and the further $g$ is from the critical value $g_c=1$. Actually, for small values of $\lambda$, this approximation turns out to be rather accurate in a significantly wider range of parameters: for example, we verified that the dynamics of $n_{k=\pi/2}$ obtained from this method for $\lambda = 0.1$, $g_0 = 8$ and $g=1.01$ agrees with the one obtained via an exact numerical
  diagonalization of the fermionic
  model for $N=20$ spins within 2\% up to time scales $t \simeq 10^3$. Beyond the time range of validity of the low-density approximation, higher-order terms in the Holstein-Primakoff transformation are eventually expected to make the system thermalize, analogously to what is experimentally observed in bosons in 1D \cite{pretherm}.
%
%

\begin{figure}
\begin{center}
\includegraphics[width=0.412\textwidth]{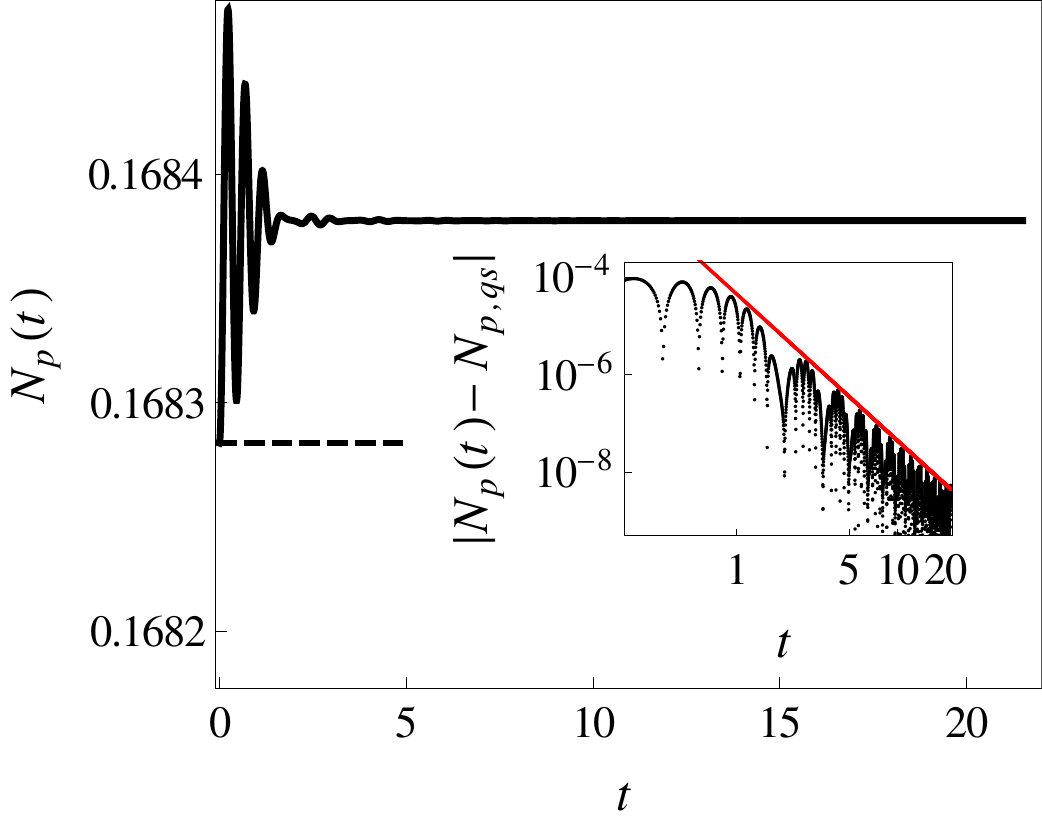}
\caption{(Color online) Time evolution of the total number of quasi-particles $N_p(t)$ 
for $N = 100$, $g_0 = 8$, $g=3.5$, and $\lambda = 0.1$. $N_p$ attains a quasi-stationary value $N_{p,qs}$ before recurrence occurs at  
$t_R \simeq 25$. 
Note that for $\lambda = 0$, $N_p$ would be constant and equal to its initial value $N_p(0) \simeq 1.68282$ (dashed line). 
For small enough $\lambda$, the difference $N_{p,qs} - N_p (0)$ as well as the amplitude of the oscillations around $N_{p,qs}$ is proportional to $\lambda N$.
The inset shows $|N_p (t) - N_{p,qs}|$ 
in double logarithmic scale and highlights the algebraic approach $\propto t^{-\alpha}$ of $N_p$ to $N_{p,qs}$; 
the straight line corresponds to $\alpha = 3$. }
\label{fig:N}
\end{center} 
\end{figure}
%
%
The main advantage of the mapping is that the diagonalization of $H$ within each sector is of polynomial complexity. Although the total number of sectors grows exponentially with $N$, as long as one can restrict the analysis to just a few of them, a numerical approach becomes effective for quite large systems, and this is generically the case for our choice: the initial state $|0\rangle_{g_0}$, in fact, contains all possible pairs of Ising quasi-particles with opposite momenta such that $I_k = 0$,  $\forall \, k$, and therefore $N_0 = N/2$. The evolution of any quantity which does not connect this particular sector with the others can be therefore computed quite easily. In particular, the operators $n_k$, $b_k$ and $b_k^\dag$ can be expressed as sums of terms oscillating in time with frequencies $\left|E_n - E_m \right|$ and $E_n + E_m$ (referred to as "slow'' and "fast'', respectively), where $\{ E_n \}_n$ is the bosonic single-particle spectrum. For small $\lambda$ the spectrum of $H_0$ is weakly perturbed, thus we can substitute $E_n$ 
with the energy $2\epsilon_n$ of a pair of quasiparticles; 
this implies that the slow frequencies range approximately from $0$ to $4$, whereas the fast ones from $4(g-1)$ to $4(g+1)$, which justifies this notion for $g>2$. 
Our numerical analysis shows that the fermion numbers $\langle \hat n_k \rangle$   
--- which are also equal to the numbers of pairs if $I_k=0$ --- display weak relaxation of the fast modes.
It is therefore more convenient to study, instead, an observable such as the total number of quasi-particles $N_p(t) = \sum_{k>0} \langle \hat n_k(t) \rangle$ which displays a marked plateau as in Fig.~\ref{fig:N}. As mentioned above, however, the dynamics of observables such as $N_p$ is characterized by a finite collection of frequencies; thus, the destructive interference which gives rise to the plateau in Fig.~\ref{fig:N} cannot last indefinitely for finite size $N$ and, in fact, we verified that oscillations start to grow again after a recurrence time $t_R \simeq N/4$.
The formalism developed here is therefore able to capture the relaxation of $N_p$ towards a \emph{pre-thermal} quasi-stationary state which, up to quantum oscillations, is approached as $\propto t^{-\alpha}$ with $\alpha \simeq 3$ (see the inset of Fig.~\ref{fig:N}). 
This same algebraic relaxation has also been observed in the average of the unperturbed Hamiltonian $\lan H_0 (g)\ran$ 
and is actually expected to characterize
every generic observable which can be expressed as a linear combination of the fermion numbers $\langle \hat n_k\rangle$ or of the number of pairs --- with possible exceptions depending on specific choices of the coefficients of these combinations. 
Furthermore, we have numerical evidence that these features are 
independent of the specific values of $N$ and $\lambda$, provided that the former is large enough and the latter small enough.
Since the dynamics at intermediate times is dictated by a quadratic integrable Hamiltonian of non-interacting bosons, a power-law approach towards the pre-thermal state is expected \cite{Barthel}.

\begin{figure}
\begin{center}
\includegraphics[trim = 50mm 180mm 50mm 35mm, clip, width=0.35\textwidth]{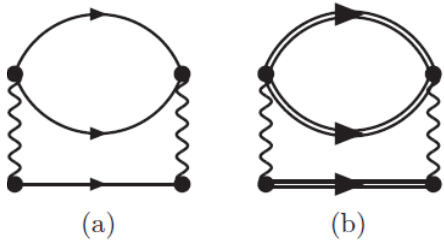}
\caption{(a) The sunset diagram. (b) The self-consistent sunset diagram. Single and double lines indicate, respectively, bare and full Green functions; wiggly lines stand for the interaction.}
\label{figsunset}
\end{center} 
\end{figure}

\emph{Diagrammatic approach.---} At longer times higher-order terms in the Holstein-Primakoff transformation cause a redistribution of the energy among the degrees of freedom of the system and are expected to lead possibly to thermalization. 
As stated above, inelastic effects can no longer be disregarded in this regime which therefore cannot be captured by our numerical approach. 
In order to investigate the mechanism which leads to relaxation and eventual thermalization in the late dynamics of this model we take below a complementary approach.
It is convenient to study first  pre-thermalization within a diagrammatic, perturbative approach at the second order in $\lambda$.
For this purpose, we employ the Dyson equation for the Green function ordered on the Keldysh contour \cite{Haug}; since tadpole diagrams do not cause relaxation \cite{Cardy}, we focus on the simplest relevant ones among the others, i.e., we assume that the self-energy is just given by the sunset diagram, see Fig.~\ref{figsunset}(a). 

%

Among the quantities one can study within this approach, the simplest is the distribution of the quasi-stationary populations \cite{MM}
\be
\label{occupy}
n_k = \lim_{T\to\infty} \frac{1}{T}\int_0^T\!\!\! \rmd t \,\langle \hat n_k  (t) \rangle =\int_{-\infty}^{\infty}\frac{\rmd\omega}{2\pi}F_k(\omega)A_k(\omega),
\ee
expressed in terms of the spectral density $A_k(\omega)$ and of the statistical function $F_k(\omega)$, where the limit is intended to hold up to the time scales of validity of perturbation theory.
For a quenched QIC ($\lambda=0$, $g\neq g_0$), $A_k(\omega)=2\pi i\delta(\omega-\epsilon_k)$ whereas $F_k$ corresponds to a Fermi-Dirac distribution function $f_k(\omega)\equiv ({\rm e}^{\beta_k\omega}+1)^{-1}$ for the GGE, with a mode-dependent inverse temperature $\beta_k=2{\rm\, th}^{-1}{(\cos \Delta_k)}/\epsilon_k$, where $\Delta_k = 2[\theta_k(g) - \theta_k(g_0)]$ \cite{Ising,Foini,Fagotti}.
This clearly highlights the role of integrability: the absence of inelastic processes prevents the QIC from redistributing the energy among the quasi-particles, and therefore thermalization as a whole.

When $V$ is turned on, $\delta(\omega-\epsilon_k)$
in $A_k$ broadens and becomes a Lorentzian with a frequency-dependent inverse life time 
$\Gamma_k(\omega) = \frac{2\lambda^2}{N^2}\sin^2(2\theta_k)\Big[f_k(\epsilon_k)\Pi(\omega+\epsilon_k)+f_k(-\epsilon_k)\Pi(-\omega-\epsilon_k)\Big]$.
$\Pi(\omega)$, calculated in second-order perturbation theory, describes effective absorption and emission of a pair of
fermions $\gamma_k$; these processes occur for  $\omega\in[2(g-1),2(g+1)]$ and $[-2(g+1),-2(g-1)]$ \cite{MM}, respectively, 
because the unperturbed spectrum $\epsilon_k$ of each fermion ranges from $g-1$ to $g+1$.
$\Gamma_k(\omega)$ quantifies the spreading of the Ising quasi-particles over the new interacting eigenmodes; close to the pronounced peak at $\omega= \epsilon_k$ (energy level shifts are disregarded here), $\Gamma_k$ determines the effective width of the Lorentzian, given by 
\begin{equation}\label{Gamma}
\Gamma_k(\epsilon_k) = \frac{\lambda^2}{4 N g} \sin^3(2\theta_k) f_k(\epsilon_k)f_k(-\epsilon_k).
\end{equation}

Analogously, $F_k(\omega)$ gets a correction to the aforementioned GGE distribution function $f_k(\omega)$, which is actually independent of $\lambda$ [see $\Gamma_k(\omega)$], $F_k(\omega)=f_k(\omega)+\frac{\lambda^2}{N^2}\sin^2(2\theta_k)\frac{f_k(-\epsilon_k)f_k(-\omega)\Pi(-\omega-\epsilon_k)-f_k(\epsilon_k)f_k(\omega)\Pi(\omega+\epsilon_k)}{2\Gamma_k(\omega)}$. This distribution function is \emph{neither} thermal \emph{nor} GGE-like and causes perturbative corrections --- characteristic of a \emph{pre-thermal} state \cite{Kollar2011} --- to appear in the occupation number $n_k$, upon integrating Eq.~\reff{occupy}. The apparent puzzle of a perturbative correction to observables, despite a pre-thermal distribution function independent of $\lambda$, 
is resolved by realizing that $F_k=f_k$ for $\omega=\epsilon_k$, hence the first 
non-vanishing correction is proportional to $\lambda^2/N$. 

\emph{Thermalization.---} Finally, we use our diagrammatic expansion
to extract physical information about thermalization dynamics in the long-time limit. In order to do so, we include in the analysis the cascade of inelastic processes triggered by the integrability breaking perturbation described by the self-consistent diagram, Fig.2(b). The resulting kinetic equation for $n_k$ admits a thermal distribution as stationary solution (it is actually the only choice which makes the collision kernel identically vanish, as it will be detailed elsewhere \cite{MM}). 
The linearization $\partial_t (\delta\hat{n}_k) = -\sum_q\hat{{\cal R}}_{kq} \delta\hat{n}_q$ of this kinetic equation for $n_k=n_k^{th.}+\delta \hat{n}_k$ around the thermal equilibrium solution $n_k^{th.}$ has a density $\rho(r)$ of relaxation rates which can be calculated by diagonalizing numerically $\hat{{\cal R}}_{kq}$. Interestingly enough, we find that $\rho(r\to0 ) \simeq r^{\zeta}$, with $\zeta\approx1/4$, and this absence
of a minimal relaxation rate is compatible with an algebraic approach to the thermal state \cite{MM}. This power-law decay might be due to the long-range nature of the interaction, as suggested by some results concerning the quench dynamics of long-range interacting models  \cite{Kastner, Long}.
Finally, under the assumptions that the quantities involved in the retarded Dyson equation vary on energy scales much larger than 
$\Gamma_k (\epsilon_k)$ and that $N$ is sufficiently large to disregard $O(N^{-2})$ corrections, we 
also calculate the quasi-particle inverse lifetime 
\begin{equation}\label{GammaSC}
\Gamma_k(\epsilon_k) = \frac{\lambda^2}{4 N g} \sin^3(2\theta_k) f^{\rm (\beta)}_k(\epsilon_k)f^{\rm (\beta)}_k(-\epsilon_k),
\end{equation}
which resembles the corresponding perturbative expression, Eq. \eqref{Gamma}, with the GGE function $f_k(\omega)$ replaced by the thermal Fermi-Dirac distribution $f^{\rm (\beta)}(\omega)$. 

\emph{Conclusions.---} We have studied the quasi-stationary and stationary states and the approach to them in a QIC weakly perturbed away from integrability. By a combination of a mapping to hard-core bosons and a perturbative analysis, we found that 
an algebraic relaxation towards the pre-thermal state emerges within an effective integrable bosonic description of the model. On the other hand, thermalization is captured by a self-consistent diagrammatic description of inelastic scattering. The absence of a minimum relaxation rate of perturbations around the thermal solution for $\hat n_k$  is consistent with the thermal state being approached in time via a power-law decay. 
Accordingly, this suggests that pre-thermalization may be accompanied, as in the present case, by a crossover between two algebraic laws.

\emph{Acknowledgments.---}We would like to thank E. Canovi for helpful comments on the numerical diagonalization of the model and P. Calabrese and M. Fabrizio for useful discussions. JM, AS and AG would like to thank the KITP for hospitality during the workshop on "Quantum Dynamics in Far from Equilibrium Thermally Isolated Systems" where part of this work has been done.

\end{document}